\documentclass[fleqn,usenatbib]{mnras}
\usepackage{siunitx}
\usepackage{physics}
\usepackage{graphicx}
\DeclareSIUnit\sr{sr}
\DeclareSIUnit\jy{Jy}
\DeclareSIUnit\erg{erg}
\title[CHIME FRB neutrino flux constraint]{Time-integrated constraint on neutrino flux of CHIME fast radio burst sources with 10-year IceCube point-source data}
\author[Luo and Zhang]{Jia-Wei Luo$^{1,2,3,4}$\thanks{ljw@hebtu.edu.cn} and Bing Zhang$^{3,4}$
\\
$^{1}$College of Physics and Hebei Key Laboratory of Photophysics Research and Application, Hebei Normal University, \protect\\ \, Shijiazhuang, Hebei 050024, China\\
$^{2}$Guo Shoujing Institute for Astronomy, Hebei Normal University, Shijiazhuang, Hebei 050024, China\\
$^{3}$Nevada Center for Astrophysics, University of Nevada, Las Vegas, NV 89154, USA\\
$^{4}$Department of Physics and Astronomy, University of Nevada, Las Vegas, NV 89154, USA\\
}

\date{Accepted XXX. Received YYY; in original form ZZZ}
\pubyear{2022}
\begin{document}
\label{firstpage}
\pagerange{\pageref{firstpage}--\pageref{lastpage}}
\maketitle

\begin{abstract}
    Despite numerous studies, the sources of IceCube cosmic neutrinos are mostly unidentified. Utilizing recently released IceCube neutrino and CHIME fast radio burst (FRB) catalogs, we examine the possibility of an association between neutrinos and CHIME/FRB catalog 1 FRBs for both the entire FRB population and individual FRBs using the unbinned maximum likelihood method. Our results do not directly support the possibility of the above-mentioned association with three weighting schemes: equal, total radio fluence, and event rate. We then attempt to constrain the diffuse muon neutrino flux upper limit from CHIME/FRB catalog 1 FRBs. After considering a completeness correction, we find the 95\% diffuse muon neutrino flux upper limit at $\SI{100}{\tera\eV}$ for all FRB sources in the universe to be $\sim\SI{1.01e-18}{\per\giga\eV\per\centi\meter\squared\per\s\per\sr}$, or $\sim 70.3\%$ of the 10-year diffuse neutrino flux observed by IceCube. Our results match the non-detection results of other studies, but we do not rule out FRBs being a significant contributor to the diffuse neutrino flux measured by IceCube.
\end{abstract}


\section{Introduction}
\label{sec:introduction}
	
The sources of diffuse cosmic neutrinos observed by the IceCube neutrino observatory \citep{icecubecollaboration2013EvidenceHighEnergyExtraterrestrial} remains a mystery \citep{meszaros2017AstrophysicalSourcesHigh,murase2019HighEnergyMultimessengerTransient,aartsen2019SearchSourcesAstrophysical,aartsen2020TimeIntegratedNeutrinoSource,liodakis2022HuntExtraterrestrialHighenergy}. 

In 2018, the IceCube collaboration reported an association between a neutrino event IceCube-170922A and the blazar TXS 0506+056 during a gamma-ray flaring state \citep{icecubecollaboration2018MultimessengerObservationsFlaring}. This has led to the speculation that blazars are the main sources of cosmic neutrinos \citep[e.g.][]{essey2010SecondaryPhotonsNeutrinos,kalashev2013PeVNeutrinosIntergalactic,kun2021CosmicNeutrinosTemporarily,das2022CosmogenicGammarayNeutrino}. However, \cite{luo2020BlazarIceCubeNeutrinoAssociation} showed that there is no compelling correlation between blazars and neutrinos in general, nor is there an association between IceCube neutrino flares and gamma-ray fluxes in Fermi-LAT monitored sources. The neutrino events clustering in the direction of TXS 0506+056 only give a significance of $2.9\sigma$, and only after including the gamma-ray flare detection at the same time did the association reach a $3.7\sigma$ significance \citep[Materials and Methods]{icecubecollaboration2018MultimessengerObservationsFlaring}. This marginal confidence level cannot firmly confirm the association between TXS 0506+056 and IceCube neutrinos, especially considering that gamma-rays are generally not associated with neutrinos.

The IceCube collaboration also reported another association between TXS 0506+056 and neutrino events in 2014/2015 discovered with an archival search \citep{icecubecollaboration2018NeutrinoEmissionDirection}. TXS 0506+056 was not at a gamma-ray flaring state at this time, and the association significance is $3.5 \sigma$. However, a refined analysis with new data and software by the IceCube collaboration revised the significance of this association down to $2.4 \sigma$ \citep{icecubecollaboration2021IceCubeDataNeutrino}, further undermining the association between TXS 0506+056 and neutrinos.

There are some neutrino event associations reported with several active galactic nuclei (AGNs) including blazars, as well as some tidal disruption events (TDEs) \citep{plavin2020ObservationalEvidenceOrigin,aartsen2020TimeIntegratedNeutrinoSource,rodrigues2021MultiwavelengthNeutrinoEmission,abbasi2021SearchMultiflareNeutrino,stein2021TidalDisruptionEvent,sahakyan2023MultimessengerStudyBlazar,icecubecollaboration2022EvidenceNeutrinoEmission}. If they are real, they are likely from particular sources with special physical conditions such as AGNs with bright Doppler-boosted jets \citep[e.g.][]{britzen2019CosmicColliderWas,plavin2020ObservationalEvidenceOrigin,reusch2022CandidateTidalDisruption}.

Recently, the IceCube collaboration reported evidence of neutrino emission from NGC 1068 \citep{icecubecollaboration2022EvidenceNeutrinoEmission}, a type 2 Seyfert galaxy. It remains to be seen whether other similar sources are also neutrino emitters. Numerous other studies have also been carried out trying to hunt down the source of the cosmic diffuse neutrino flux, so far to no avail \citep[e.g.][]{adrian-martinez2012SearchNeutrinoEmission,hooper2019ActiveGalacticNuclei,capel2020BayesianConstraintsAstrophysical,abbasi2021ModelindependentAnalysisNeutrino,stathopoulos2021ProbingNeutrinoEmission,goswami2021SearchHighenergyNeutrino,zhou2021SearchHighenergyNeutrino,theicecubecollaboration2021NewSearchNeutrino,abbasi2022SearchHighEnergyNeutrino,abbasi2022SearchAstrophysicalNeutrinos,abbasi2022SearchingHighEnergyNeutrino,abbasi2022SearchesNeutrinosGammaRay,li2022InvestigatingCorrelationsIceCube,kheirandish2023DetectingHighenergyNeutrino,abbasi2023IceCubeSearchNeutrinos,abbasi2024AllskySearchTransient}. In particular, other than some individual sources, none of the classes of cosmic objects have been firmly found to be associated with neutrinos at higher than a $5\sigma$ confidence level.

The IceCube collaboration also reported detection of high-energy neutrinos from the Galactic plane
\citep{icecubecollaboration2023ObservationHighenergyNeutrinos,abbasi2023SearchExtendedSources}, accounting for $\sim\SIrange{6}{13}{\percent}$ of the astrophysical neutrino flux at $\SI{30}{\tera\eV}$. The IceCube collaboration found the most significant source candidate to be 3HWC J1951+266 at a $2.6\sigma$ confidence level. The exact source of this galactic neutrino flux remains to be pinpointed.

Fast radio bursts (FRBs) are short cosmic radio transients whose origins are yet to be fully uncovered \citep{lorimer2007BrightMillisecondRadio,thornton2013PopulationFastRadio,platts2019LivingTheoryCatalogue,petroff2019FastRadioBursts,cordes2019FastRadioBursts,zhang2020PhysicalMechanismsFast,xiao2022FastRadioBursts,petroff2022FastRadioBursts,zhang2023PhysicsFastRadio}, and are thought to be possible neutrino emitters \citep{metzger2020NeutrinoCounterpartsFast,qu2022NeutrinoEmissionFast}. Some early attempts searching for possible associations between FRBs and IceCube neutrinos have been carried out \citep{fahey2017SearchNeutrinosFast,aartsen2018SearchNeutrinoEmission,albert2019SearchHighenergyNeutrinos,aartsen2020SearchMeVTeV,nicastro2021MultiwavelengthObservationsFast,abbasi2023SearchCoincidentNeutrino}, but no significant results have been obtained yet.

Recently, the Canadian Hydrogen Intensity Mapping Experiment (CHIME)/FRB Collaboration released the first catalog, containing a total of 492 FRB sources, among which 18 are repeating sources  \citep{chime/frbcollaboration2021FirstCHIMEFRB}. The recently released 2008--2018 all-sky point-source IceCube data \citep{icecubecollaboration2021IceCubeDataNeutrino}, on the other hand, have 1134450 neutrino events spanning ten years. The availability of these two large catalogs makes it possible to perform a direct search of associations between the two types of astrophysical events. Note that the observing times of the two catalogs do not overlap. However, some FRB repeaters (e.g. FRB 121102) have been known to persist for at least a decade and some recently detected repeaters do not show a significant evolution of burst rate on a timescale of years. Even for apparent non-repeating FRBs, rate arguments suggest that they likely do not originate from cataclysmic events \citep{ravi2019PrevalenceRepeatingFast,luo2020FRBLuminosityFunction} and the lack of multiple detection may be a result of low repetition rate or that most repeated bursts are below the radio detection sensitivity threshold. It is very likely that these ``non-repeating" FRBs also emitted additional bursts in the past, but these other bursts were not detected.
It is therefore reasonable to assume that the activity level of FRB sources remained nearly constant during the timescale difference between the two catalogs.

In this work, we conduct a search for association between FRBs and neutrinos for both individual FRBs and the CHIME FRB population. While we did not find any significant association between IceCube neutrinos and any FRB or the FRB population, we constrain the contribution from CHIME FRBs to the IceCube diffuse neutrino flux.

\section{Unbinned likelihood analysis of individual FRBs}
\label{sec:frb_individual}

We employ the commonly used unbinned likelihood method \citep{braun2008MethodsPointSource,braun2010TimedependentPointSource} to search for associations of neutrino events with individual FRBs.

For a single source, the likelihood function of the unbinned method is written as:
\begin{equation}
    \mathcal{L}=\prod_{i=1}^N\qty[\frac{n_s}{N}S_i+\qty(1-\frac{n_s}{N}B_i)],
\end{equation}
where $i$ is the index of the neutrino event, $N$ is the total number of neutrino events, $S_i$ is the probability density function (PDF) of the source associated events, $B_i$ is the PDF of background events, $n_s$ is the number of signal neutrino events emitted by the source. $n_s$ is a free variable and should be determined by maximizing the likelihood function $\mathcal{L}$ while noting that $n_s\ge0$.

The source PDF is defined as:
\begin{equation}
    S_i=\frac{1}{2\upi\sigma_i^2} \exp(-\frac{\theta_i^2}{2\sigma_i^2}),
\end{equation}
where $\sigma_i$ is the angular localization error of each neutrino event, $\theta_i$ is the angular separation between the hypothetical source and each neutrino event $i$. We do not include any time component, as the CHIME and IceCube catalogs do not have time overlap.

The background PDF is defined as:
\begin{equation}
B_i=\frac{\mathrm{Number\;of\;neutrino\;events\;in\;dec\;ring}}{\mathrm{Area\;of\;dec\;ring \times Total\;number\;of\;neutrino\;events}},
\end{equation}
where the dec ring is the ring-shaped region $\pm 3^{\circ}$ in declination around the $i$th neutrino event. For neutrino events within $3^{\circ}$ of the poles, we use the values at $\pm 87^{\circ}$ correspondingly. The area of the dec ring is its angular area in units of steradian.

After finding the expected $n_s$ value, $\hat{n}_s$ by maximizing $\mathcal{L}$, we can then define a test statistic:
\begin{equation}
    \mathrm{TS}=2\log\frac{\mathcal{L}(\hat{n}_s)}{\mathcal{L}(0)}.
\end{equation}
This test statistic is a likelihood ratio of two hypothesises:
$H_1$: Of all the $N$ neutrino events, $\hat{n}_s$ events come from the hypothetical source.
$H_0$: All of the $N$ events are background events.
A high TS value indicates that the null hypothesis $H_0$ can be rejected and that at least some neutrino events are associated with the individual source. 

\begin{figure}
	\includegraphics[width=0.49\textwidth]{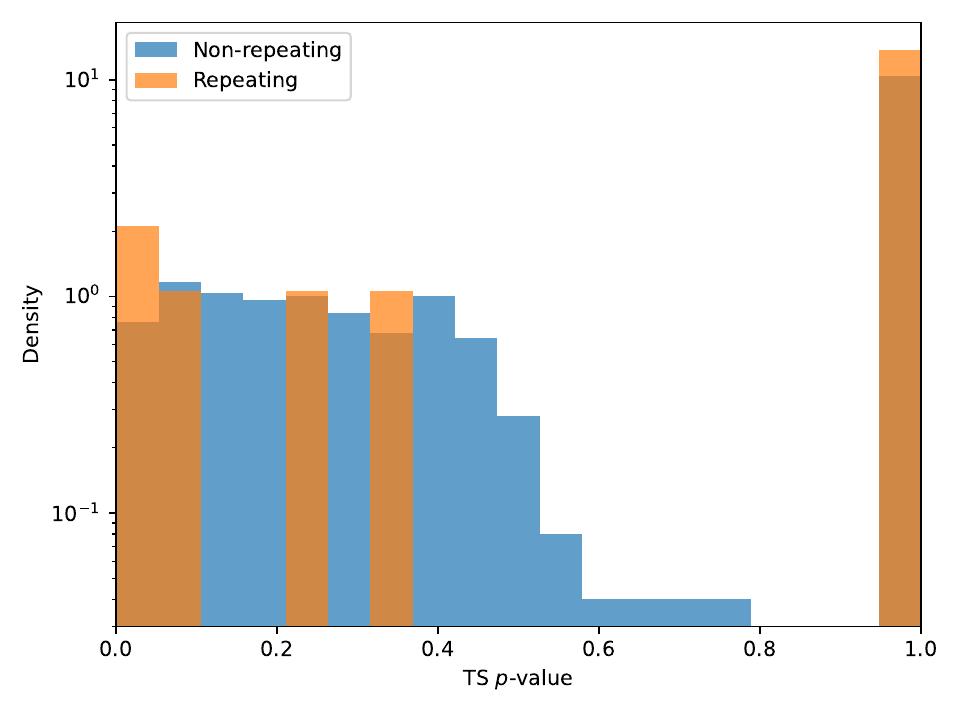}
	\caption{Distributions of $p$-values for CHIME repeating and non-repeating FRBs. The peak at $p=1.0$ corresponds to $TS=0$.}
	\label{fig:pvalues}
\end{figure}

\begin{figure}
	\includegraphics[width=0.49\textwidth]{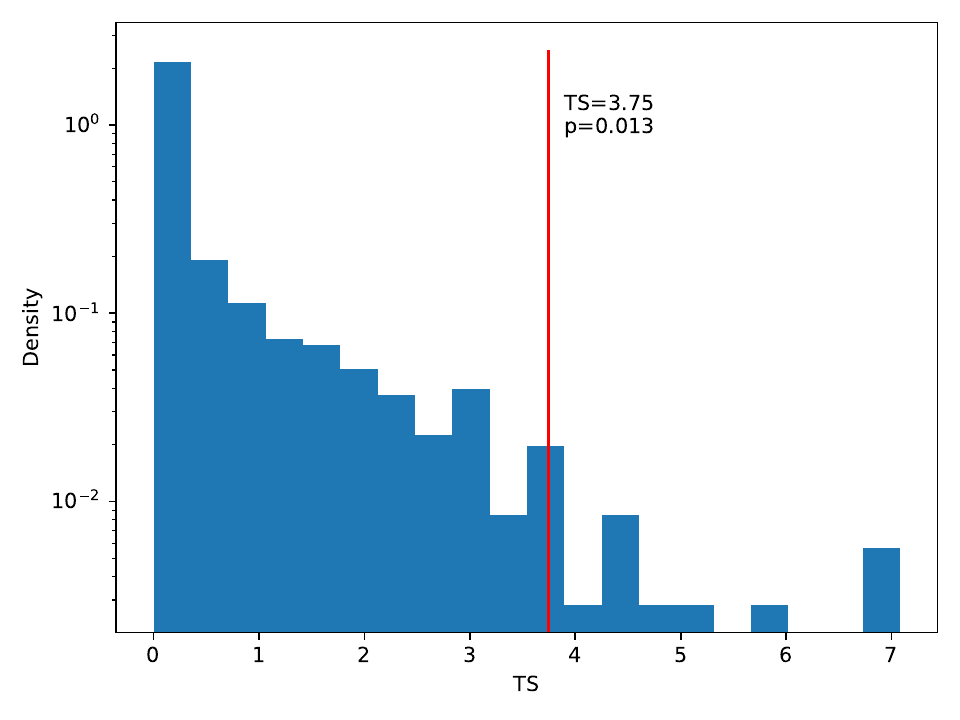}
	\includegraphics[width=0.49\textwidth]{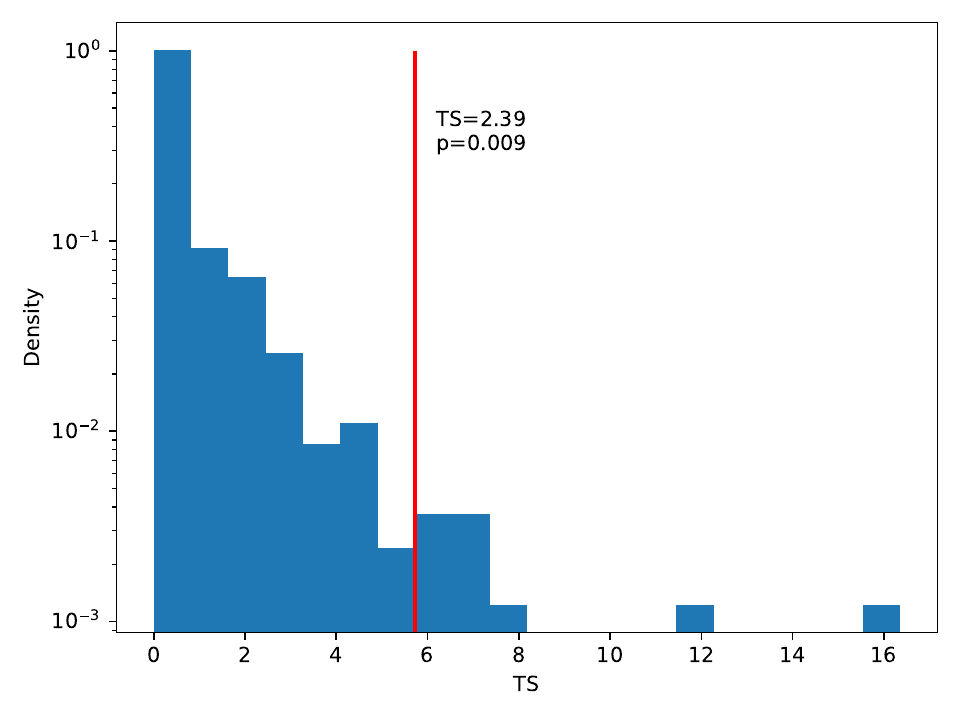}
	\caption{Distributions of simulated TS values for repeating and non-repeating FRBs with lowest $p$-values. The observed TS are marked with red lines. Upper: Repeater FRB20180908B, Lower: Non-repeater FRB20190502B.}
	\label{fig:ts_simulation}
\end{figure}

As the local neutrino background distributions of the FRBs are different, we cannot directly compare the $n_s$ and $TS$ values across the FRBs. In order to better test the significance of the possible association between CHIME FRBs and IceCube neutrinos, we conduct a Monte Carlo simulation for each of the 220 FRBs with fitted $n_s$ and TS values greater than 0. We randomly reshuffle all neutrino events in the RA direction, and calculate the new TS values with the same method described above. Because IceCube is located at the south pole, its observed neutrino events are isotropic in the RA direction. However, IceCube neutrinos do have an anisotropy in the declination direction. Coupled with the distribution of CHIME FRBs, this can result in false detection. Therefore, We only reshuffle neutrino events in the RA direction, so that we can keep their number distributions in the dec direction.

We repeat this Monte Carlo simulation 1000 times for each FRB, and we compare the observed TS values with the simulated TS value distributions. A $p$-value can be subsequently obtained by calculating the percentile that the observed values lie in. The distributions of $p$-values are shown in Figure \ref{fig:pvalues}. There is a large peak at $p=1.0$ corresponding to $TS=0$, showing an under-fluctuation. The repeating and non-repeating FRBs with the lowest $p$-values are FRB20180908B and FRB20190502B, respectively. The simulated TS distributions of the two FRBs are shown in Figure \ref{fig:ts_simulation}\footnote{We use density rather than counts per bin as vertical axes for Figure \ref{fig:pvalues} to avoid non-uniformity between repeaters and non-repeaters. }. The $p$-values of the two FRBs are 0.013 and 0.009, corresponding to significance levels of $2.23\sigma$ and $2.37\sigma$, respectively. For the lowest $p$-value of 0.009, we consider the total number of the FRBs tested, 492. We find a post-trial $p$-value of $1 - (0.991)^{492}=0.99$. This is a very significant under-fluctuation. Our results are therefore entirely consistent with background expectations.

There are 21 FRBs with $p$-values lower than 0.05, among them 19 non-repeating and 2 repeating. However, because of the look-elsewhere effect, this number of FRBs showing possible associations with neutrinos at this confidence level should not be interpreted as evidence for associations, since 5\% of the FRB sample size of 492 is 24.6. This test result is therefore an under-fluctuation relative to background expectations, and our results are therefore fully consistent with the null hypothesis of no neutrino--CHIME FRB correlation.

\section{Stacked unbinned likelihood analysis of CHIME FRB population}

The likelihood functions of individual sources can also be combined to form a stacked likelihood function by modifying the source PDF \citep{abbasi2011TIMEINTEGRATEDSEARCHESPOINTLIKE,aartsen2017CONTRIBUTIONFERMI2LACBLAZARS,smith2021RevisitingAGNSource,abbasi2022SearchAstrophysicalNeutrinos}
\begin{equation}
    S_i=\frac{\sum_{j=1}^M w_jS_{ij}}{\sum_{j=1}^M w_j},
\end{equation}
where $M$ is the total number of possible neutrino sources (in this study FRBs), $w_j$ is a weight factor denoting the relative neutrino flux from each source, and
\begin{equation}
    S_{ij}=\frac{1}{2\upi\sigma_i^2} \exp(-\frac{\theta_{ij}^2}{2\sigma_i^2}).
\end{equation}
Since the neutrino event distribution is the same for all possible sources, the background PDF $B_i$ can be carried over without any modification.

We utilize three weighting schemes:
\begin{itemize}

    \item{Equal}
    \begin{equation}
        w_j=1,
    \end{equation}
    where all FRBs carry the same weight.
    
    \item{Total radio fluence}
    \begin{equation}
        w_j=\sum_k F_{\nu k},
    \end{equation}
    where $F_{\nu k}$ is the observed band-average specific radio fluence of the $k$th burst of the FRB given by the CHIME/FRB catalog 1 in units of \si{\jy\milli\s}. For non-repeaters, there is only one burst and no summation is needed.

    We test the hypothesis that the neutrino flux from FRBs scales with burst radio fluence for this case. We do not need to normalize the weights here, because we included the normalization in the formula for $S_i$. Since FRBs are bursts, in contrast with continuous emitters like blazars, their neutrino fluxes may also depend on their durations. Therefore we use total radio fluence instead of flux.
    
    \item{Event rate}
    \begin{equation}
        w_j=\begin{cases}
		    1, & \text{Non-repeaters}\\
            \text{Observed number of bursts}, & \text{Repeaters}
		    \end{cases}.
    \end{equation}
    Being bursts, FRBs' neutrino fluxes may depend on their event rates. We use the observed number of bursts as an estimation of the event rate of repeating FRBs. The weights for non-repeating FRBs are taken to be 1 in this case.
    
\end{itemize}

We test three FRB samples with the three weighting schemes, non-repeating FRBs, repeating FRBs, and all FRBs. The fitted TS for all 9 combinations is 0.

\begin{figure}
\includegraphics[width=0.49\textwidth]{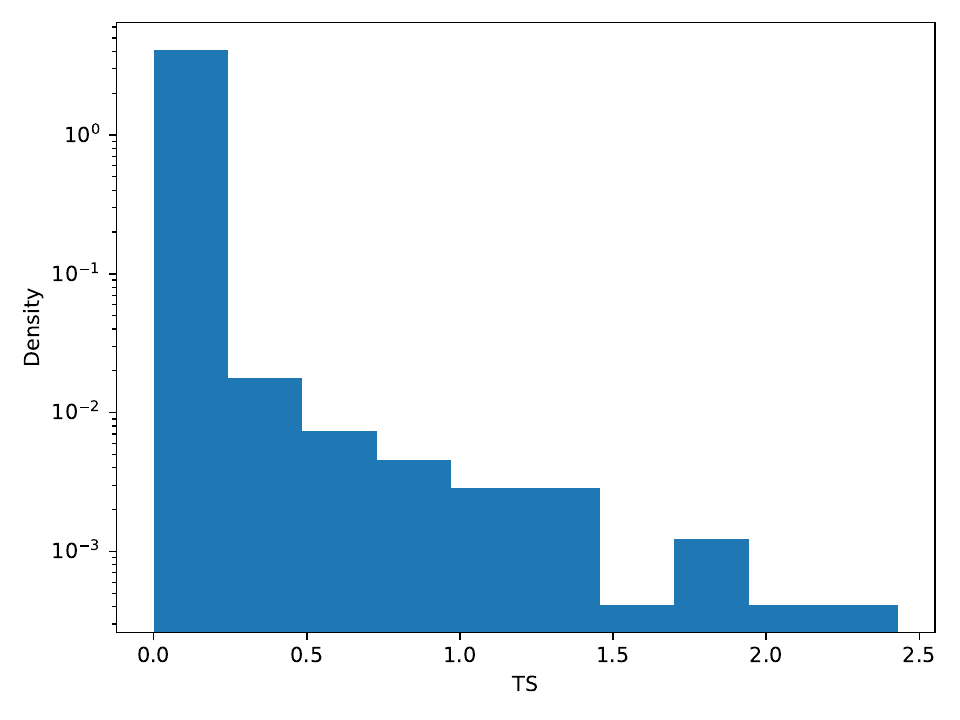}
\caption{Distribution of simulated TS values with all FRBs and the equal weighting scheme.}
\label{fig:stack_test}
\end{figure}

In order to show that the non-detection of FRB -- neutrino association in the stacked analysis is not a result of errors in our method, we scramble the neutrino data in the RA direction similar to the methods described in Section \ref{sec:frb_individual} and calculate TS values on the scrambled data. The distribution of simulated TS values with all FRBs and the equal weighting scheme is shown in Figure \ref{fig:stack_test}. Among 10000 tests, only 4\% returned TS values greater than 0. Therefore our results disfavor the possibility of the entire population of CHIME FRBs being significant neutrino sources.

We then move on to determine the upper limit of the CHIME FRBs' contribution to the IceCube diffuse neutrino flux.

The number of observed neutrino events from a single source with index $j$ can be estimated with:
\begin{equation}
    n_{sj}=w_j \sum_{k}T_k\int_{E_{min}}^{E_{max}}\mathrm{A_{eff,k}}(E_{\nu},\delta)\Phi_j(E_{\nu})\;\dd E_{\nu},
    \label{eq:constrain_ns}
\end{equation}
where $k$ is the year of observation of the neutrino catalog. The sum over $k$ is needed as the observing time and detector effective area are different in different years. $T_k$ is the total observing time of the detector at the $k$th year of the catalog, $\mathrm{A_{eff,k}}$ is the effective area of the detector at the $k$th year of the catalog, which is a function of neutrino energy $E_{\nu}$ and declination $\delta$ of the source, $\Phi_j$ is the neutrino spectrum of the source, $E_{min}$ and $E_{max}$ are the minimum and maximum energies of the detector band, and $w_j$ is the source weight.

For a simple estimation, we can assume the neutrino spectrum follows a power law:
\begin{equation}
    \Phi_j = \Phi_0 \qty(\frac{E_{\nu}}{E_0})^{-\alpha}\times\SI{e-20}{\per\giga\eV\per\centi\meter\squared\per\s\per\sr},
\end{equation}
where $\Phi_0$ is a dimensionless normalization constant, $E_0$ is a normalization energy, in this study we adopt $E_0=\SI{100}{\tera\eV}$, $\alpha$ is the powerlaw index. $\SI{e-20}{\per\giga\eV\per\centi\meter\squared\per\s\per\sr}$ is another normalization constant we add to normalize the units, as well as to make the $\Phi_0$ constants for the power law and physical models comparable.

A more sophisticated estimation of the neutrino spectrum from FRBs is given by \citet{qu2022NeutrinoEmissionFast}. The neutrino flux can be similarly written as:
\begin{equation}
    \Phi_j = \Phi_0 \Phi_q\qty(E_{\nu}),
\end{equation}
where $\Phi_q\qty(E_{\nu})$ is the neutrino spectrum given by \citet[Figure 6]{qu2022NeutrinoEmissionFast}.

For the simple powerlaw model, the total neutrino flux from a population of sources can be calculated as the sum of individual fluxes, $n_s=\sum_{j=1}^M n_{sj}$, $\Phi=\sum_{j=1}^M w_j\Phi_j/(4\pi)$. The $4\pi$ factor is needed to convert the sum of individual FRBs' contributions to the total neutrino flux to the total diffuse flux.

However, we also need to consider that the above-mentioned models are based on the average of the diffuse neutrino flux from all FRBs, but the CHIME FRB catalog only contains a small fraction of them. This is especially evident considering that CHIME can only observe a finite area of the sky during a given time, while IceCube can monitor neutrino events across all the sky. Also, the CHIME FRB catalog only contains FRB sources detected in one year of observation, while the IceCube catalog we use contains neutrino events in ten years' observation time. Therefore, we need a catalog completeness factor in the flux to account for the difference in FRB population in the CHIME catalog and the physical model. The total diffuse neutrino flux for the physical model can be written as:
\begin{equation}
    \Phi=\frac{1}{C}\sum_{j=1}^M \frac{w_j\Phi_j}{4\pi},
\end{equation}
where $C$ is the catalog completeness factor.

In considering the completeness correction, we generally follow \citet{ai2021TrueFractionsRepeating} and make two assumptions: 1. The observed repeating and non-repeating FRBs are all repeating, only with different rates of repetition; 2. The total number of such repeating FRB sources remains roughly constant in the Universe. Then, following \citet{qu2022NeutrinoEmissionFast}, we further make two assumptions: 1. The neutrino emission of a FRB source is transient in a ``microscopic" view (i.e. neutrino flux increases significantly during bursts), but the ``macroscopic" mean neutrino flux is roughly proportional to the burst rate; 2. There are many sub-threshold bursts undetectable by CHIME, but the neutrinos from those bursts can be detected by IceCube. This is evident from the significant difference in the number of detected bursts by CHIME and FAST on the same repeating FRB source. \citep[e.g.][]{li2021BimodalBurstEnergy,xu2022FastRadioBurst}.

To estimate the catalog completeness factor, we consider the CHIME all-sky FRB rate of around $\sim820$ per day \citep{chime/frbcollaboration2021FirstCHIMEFRB}, and the total number of bursts in the CHIME FRB catalog including repeated bursts, 536. The total observation time contained in the CHIME catalog is 342 days \citep{chime/frbcollaboration2021FirstCHIMEFRB}. Then, the catalog completeness factor can be estimated with $C\approx\frac{536}{820\times342}\approx\num{1.91e-3}$. Note that our completeness correction factor only accounts for missed FRBs observable by CHIME without considering the true FRB rate that depends on luminosity function and redshift distribution \citep[e.g.][]{zhang2022CHIMEFastRadio,shin2023InferringEnergyDistance,lin2024RevisedConstraintsFast,zhang2024RevisitingEnergyDistribution}. 

A more accurate way to determine the completeness factor would be to model the total FRB flux considering the luminosity function and redshift evolution and then compare with the observed FRB flux. However, since the redshift evolution of FRBs is not fully understood currently \citep[e.g.][]{zhang2022CHIMEFastRadio,shin2023InferringEnergyDistance,zhang2024RevisitingEnergyDistribution}, we adopt our simplistic model. The true completeness factor could be much lower and the true upper limit much higher if there are many more distant and faint FRBs not included in the all-sky FRB rate.

Because the neutrino catalog we are comparing against is also the observed sample without considering luminosity function and redshift distribution, we believe our treatment is more relevant to the problem.

Based on Table VI of \citet{feldman1998UnifiedApproachClassical}, we can find the $n_s$ value corresponding to a 95\% upper limit when the main search did not find any association, $n_s=3.09$. Then, by altering the normalization constant $\Phi_0$ in the neutrino spectrum and observing the resulting flux $\Phi$ and $n_s$ values until $n_s=3.09$, we can obtain a 95\% confidence level for neutrino flux emitted from FRBs.

\begin{figure}
	\includegraphics[width=0.49\textwidth]{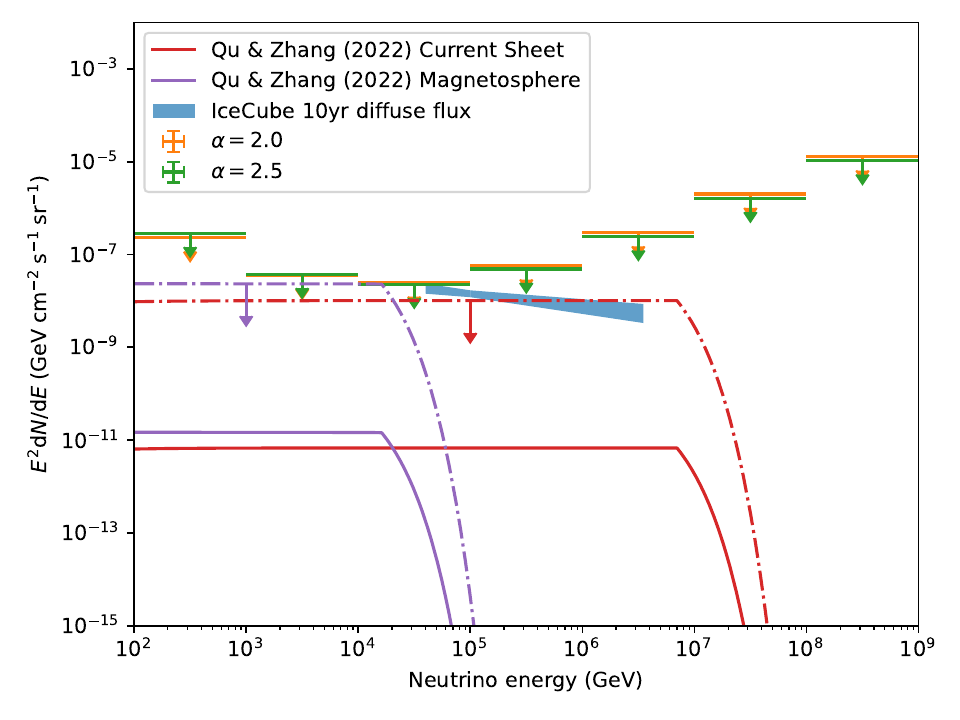}
	\caption{95\% upper limit for diffuse muon neutrino flux from FRBs in the powerlaw and physical spectrum models after completeness correction. The 10-year observed IceCube diffuse muon neutrino flux \citep{stettner2019MeasurementDiffuseAstrophysical} is shown as a blue region. For physical models, the predicted and the upper limit fluxes are drawn in the same color and cover roughly the same energy range, with downward arrows marking the upper limits and dash-dotted lines marking the predicted neutrino fluxes}.
	\label{fig:stacked}
\end{figure}

In Figure \ref{fig:stacked}, we show the 95\% diffuse flux upper limit of three types of spectral models: simple powerlaw with $\alpha=$ 2.0 and 2.5, and the physical spectra given by \citet{qu2022NeutrinoEmissionFast} after completeness correction. For the powerlaw model, we uniformly divide the IceCube energy range into 7 bins so that each bin represents one order of magnitude. We assume the calculated $n_s$ comes from each bin to calculate the diffuse neutrino flux upper limits correspondingly. We do not divide the energy range into bins for the physical models. We find the upper limits from different weighting schemes do not differ significantly, therefore we only show the results from equal weighting. We also show the predicted FRB diffuse muon neutrino flux by \citet{qu2022NeutrinoEmissionFast}, as well as the 10-year observed IceCube diffuse muon neutrino flux \citep{stettner2019MeasurementDiffuseAstrophysical} in Figure \ref{fig:stacked}.

As shown in Figure \ref{fig:stacked}, after the completeness correction, the calculated 95\% diffuse fluxes are near the 10-year observed IceCube diffuse neutrino flux \citep{stettner2019MeasurementDiffuseAstrophysical}, and $\sim1600\times$ higher than the physical models presented by \citet{qu2022NeutrinoEmissionFast}. The true neutrino flux from FRBs, if there is any, could be substantially lower than the calculated upper limits. Based on the upper limit from the current sheet model after the completeness correction, we can constrain the 95\% upper limit of the contribution of all FRBs to the diffuse muon neutrino flux at $\SI{100}{\tera\eV}$ to be $\sim\SI{1.01e-18}{\per\giga\eV\per\centi\meter\squared\per\s\per\sr}$, or $\sim 70.3\%$ of the observed IceCube diffuse muon neutrino flux. The upper limits are fairly close to the observed IceCube diffuse muon neutrino flux. This shows that FRBs are still allowed to be a significant contributor to the diffuse neutrino flux.


\section{Conclusions and discussions}
\label{sec:discussion}

In this paper, we perform a search for possible associations between CHIME/FRB catalog 1 FRB sources and IceCube neutrinos with both individual FRBs and the population of FRB sources with temporally non-overlapping data. Our result does not show any significant association. 

Then, with a stacking analysis method, we constrain the 95\% upper limit of the contribution to the IceCube diffuse muon neutrino flux from CHIME FRBs. Based on the upper limit from the current sheet model after completeness correction, the FRB muon neutrino flux upper limit is $\sim\SI{1.01e-18}{\per\giga\eV\per\centi\meter\squared\per\s\per\sr}$ at $\SI{100}{\tera\eV}$, or about $70.3\%$ of the 10-year observed IceCube diffuse muon neutrino flux.

The non-detection of significant IceCube neutrino---CHIME FRB association matches the non-detection of neutrinos from FRB directions by other studies 
\citep{fahey2017SearchNeutrinosFast,aartsen2018SearchNeutrinoEmission, albert2019SearchHighenergyNeutrinos,aartsen2020SearchMeVTeV,nicastro2021MultiwavelengthObservationsFast,abbasi2023SearchCoincidentNeutrino}. This should not come as a surprise as physical models typically predict neutrino fluxes from FRBs to be much lower than the sensitivity of current neutrino detectors. Furthermore, the stacking analysis does not rule out the possibility of CHIME FRBs being major neutrino emitters. A more comprehensive FRB catalog and a neutrino catalog with longer observation time are needed to provide a tighter constraint on the physical models for neutrino emission from FRBs.

The detection of FRB 200425 from the Galactic magnetar SGR J1935+2154 \citep{chime/frbcollaboration2020BrightMilliseconddurationRadio,bochenek2020FastRadioBurst,li2021HXMTIdentificationNonthermal,mereghetti2020INTEGRALDiscoveryBurst,ridnaia2021PeculiarHardXray,tavani2021XrayBurstMagnetar} places magnetars as the leading candidate source of FRBs. It is possible that all FRBs originate from magnetars, with the apparently non-repeating ones being repeaters with very long waiting times between bursts \citep[e.g.][]{lu2020UnifiedPictureGalactic,zhang2020PhysicalMechanismsFast}. If so, neutrinos may be continuously generated from the FRB sources.

In particular, \cite{zhang2003HighEnergyNeutrinosMagnetars} have shown that in steady state magnetars are able to make neutrinos with a few TeV characteristic energy through photomeson interaction between accelerated protons and surface X-ray photons. \cite{murase2009ProbingBirthFast} showed that new-born magnetars could produce significant high-energy neutrinos through hadronic interactions between accelerated protons and the surrounding stellar ejecta. However, these two emission components either have too low a diffuse flux \citep{zhang2003HighEnergyNeutrinosMagnetars} or too high a neutrino energy \citep{murase2009ProbingBirthFast} for IceCube to detect. 
	
Alternatively, neutrinos may be produced during X-ray flaring processes, some of which are associated with FRBs \citep{lin2020NoPulsedRadio}. \cite{metzger2020NeutrinoCounterpartsFast} estimated the neutrino flux within the framework of the synchrotron maser model invoking relativistic shocks and found that the predicted neutrino fluxes are negligibly low. \cite{qu2022NeutrinoEmissionFast}, on the other hand, studied photomeson interaction in the near zone regions, either within the magnetosphere or in the current sheet region just outside the magnetosphere. They found that the neutrino flux could be much stronger than the estimation by \cite{metzger2020NeutrinoCounterpartsFast}. However, the predicted level is still about 1-2 orders of magnitude below the upper limit we placed. This suggests the need to continuously search for FRB-neutrino associations with ever more sensitive neutrino detectors. Since the predicted neutrino flux from FRBs depends on the particle acceleration site with the magnetospheric model having the highest fluxes \citep{qu2022NeutrinoEmissionFast}, future detection of neutrinos from FRB sources would lend indirect support of the magnetospheric origin of FRB emission from magnetars \citep[e.g.][]{yang2018BunchingCoherentCurvature,wang2019TimeFrequencyDownward,wadiasingh2019RepeatingFastRadio,lu2020UnifiedPictureGalactic,wadiasingh2020FastRadioBurst,yang2021FastRadioBursts,zhang2022CoherentInverseCompton,wang2022MagnetosphericCurvatureRadiation}.

\section{Acknowledgements}
\label{sec:acknowledgements}

This work is partially supported by the Top Tier Doctoral Graduate Research Assistantship (TTDGRA) and Nevada Center for Astrophysics at the University of Nevada, Las Vegas. JW Luo is supported by Hebei Natural Science Foundation (grant no. A2024205004), Science Research Project of Hebei Education Department (grant no. QN2024287) and Science Foundation of Hebei Normal University (grant no. L2024B07). We thank the referee Justin Vandenbroucke for many helpful suggestions that greatly improved this paper. The authors thank Ali Kheirandish for discussions and very helpful suggestions on this study. We also thank Jeremy Heyl, Clancy James, Claudio Kopper and Timothy Linden for pointing out the issues in the analysis of our earlier study on this subject.

\section*{Data availability}
\label{sec:data_availability}
The data used in this paper are public. The CHIME/FRB catalog is available at \url{https://www.chime-frb.ca/catalog}, and the IceCube 10-year all-sky point-source data is available at \url{https://icecube.wisc.edu/data-releases/2021/01/all-sky-point-source-icecube-data-years-2008-2018/}. The code used in this paper can be shared upon request to the authors.

\bibliographystyle{mnras}
\bibliography{frb-neutrino}
\bsp
\label{lastpage}
\end{document}